\documentstyle[aps,preprint]{revtex}
\begin{document}
\draft
\title{Explicit, analytical solution of scaling quantum graphs}
\date{\today}
\author{Yu. Dabaghian and R. Bl\"umel}
\address{Department of Physics, Wesleyan University,
Middletown, CT 06459-0155, USA\\
E-mail: ydabaghian@wesleyan.edu}
\maketitle

\begin{abstract}
Based on earlier work on regular quantum graphs we show
that a large class of scaling quantum graphs with
arbitrary topology
are explicitly analytically solvable.
This is surprising since quantum graphs
are excellent
models of quantum chaos and quantum chaotic systems are
not usually explicitly analytically solvable.
\end{abstract}
\maketitle

Integrable systems are rare. Many have become textbook classics, such as the harmonic 
oscillator and the hydrogen atom. It is therefore always a noteworthy event when another 
integrable system is discovered. The purpose of this Letter is to add an important class
of quantum systems to this set: quantum graphs. Quantum graphs have a long history in 
mathematics, chemistry and physics (see, e.g. the excellent review by P. Kuchment\cite{Kuch}).
They have recently come to the attention of many researchers as models of quantum chaos 
\cite{QGT1,QGT2,KS,SQA}. It may therefore be surprising that a large and important
subset of quantum graphs, scaling quantum graphs, is explicitly integrable in the
form $E_n=f(n)$, where $n$ is an integer, $E_n$ are the energy levels of the graph and $f$ 
is a function that can be constructed explicitly.
Because of the quantum chaos connection it already caused quite a stir when about a year
ago explicit solutions of a sub-class of quantum graphs, regular quantum graphs, were 
discovered \cite{JETPL,PRL}. This is so, because for any given $n$ the energy level $E_n$ 
is individually computable, without knowledge of the other quantum levels. Thus the counting 
index $n$ is a quantum number which is generally believed not to exist in a quantum system 
such as quantum graphs that show so many features of quantum chaos \cite{Gutz,STOECK}. While 
in the case of regular quantum graphs one might shrug off the existence of explicit solutions 
by pointing to the special, non-generic nature of regular quantum graphs (only simply connected, 
linear and circular graphs have so far been identified as members of the set of regular quantum 
graphs), this is now no longer possible. Thus the results reported in this Letter are a 
substantial, qualitative step forward with profound implications for quantum graph theory and
quantum chaos. Focussing on scaling quantum graphs \cite{JETPL,PRL} excluding graphs with
tunneling and bound states for convenience (tunneling and bound states are discussed in \cite{FPRE}),
we will prove below that these quantum graphs, no matter how complex their topology, are
explicitly solvable analytically. We emphasize that {\it scaling} quantum graphs \cite{JETPL,PRL} 
represent a huge class of graphs, far larger than and {\it including} the standard, undressed
graphs primarily studied in the literature \cite{Kuch,QGT1,QGT2,KS,SQA}. The class of quantum 
graphs for which we prove the existence of explicit solutions comprise quantum graphs with 
arbitrary topology, arbitrary scaling potentials on their bonds and arbitrary scaling $\delta$-functions 
on their vertices.

It has been shown before \cite{JETPL,PRL} that the spectral function of scaling quantum graphs is 
given by
\begin{equation}
   g^{(0)}(k)=
   \cos(S_0 k+\varphi_0) -
   \sum_{j=1}^N\, a_j\,
   \cos(S_jk+\varphi_j),
 \label{1}
\end{equation}
where $S_0$ is the action length of the graph, $S_j<S_0$ are certain combinations of the graph actions, 
$N$ is the number of different graph action combinations in (\ref{1}), $\varphi_0$, $\varphi_j$ are constant 
phases and $a_j$ are constant amplitudes. The roots $k^{(0)}_n=\sqrt{E_n}$ of the spectral equation
\begin{equation}
 g^{(0)}(k^{(0)}_n)=0
\label{1p}
\end{equation}
are the wave numbers of the eigenenergies $E_n$ of the graph. It was shown in \cite{PRL} and put on a 
solid mathematical foundation in \cite{MF} that (\ref{1p}) is explicitly solvable in the form $k^{(0)}_n=\ldots$ 
for regular quantum graphs \cite{JETPL,PRL,MF} which satisfy
\begin{equation}
   \sum_{j=1}^N\,
   |a_j|<1.
 \label{2}
\end{equation}
The vast majority of quantum graphs does not satisfy the regularity condition (\ref{2}).
Up to now this was believed to be the hedge that makes general quantum graphs resilient against explicit
solution and thus preserves one of the central tenets of quantum chaos theory, i.e. the loss of quantum
numbers \cite{Gutz,STOECK}. However, there is a way to solve all scaling quantum graphs, even if they don't 
fulfill (\ref{2}), by constructing a chain of spectral functions terminating with a function that does satisfy 
(\ref{2}), and then making full use of (\ref{2}) and the associated mathematical machinery developed for
regular quantum graphs \cite{MF}. Thus the theory of regular quantum graphs provides us with a powerful
point of departure for solving the more general problem of all scaling quantum graphs.

Taking the $m$'th derivative of (\ref{1}) and dividing by $S_0^m$ we obtain
\begin{equation}
   g^{(m)}(k)=
   \cos(S_0 k+\varphi_0+m\pi/2) -
   \sum_{j=1}^N\, b_j^{(m)}\,
   \cos(S_jk+\varphi_j
    +m\pi/2),
 \label{3}
\end{equation}
where $b_j^{(m)}=a_j(S_j/S_0)^m$. Since $S_j<S_0$, there always exists an $M$ such that $\sum |b_j^{(M)}|<1$,
i.e. for $m=M$ (\ref{3}) satisfies the regularity condition (\ref{2}) and the roots $k_q^{(M)}$ of
$g^{(M)}(k)=0$ are explicitly computable via explicit analytical, convergent periodic-orbit expansions as
shown in \cite{JETPL,PRL} and proved rigorously in \cite{MF}.

Having bootstrapped the roots of (\ref{1}) at the level $M$ by taking the $M$'th derivative of (\ref{1}), 
we now have to go backwards to the levels $M-1$, $M-2$, $\ldots$, until we reach level 0 and possess $k_n^{(0)}$ 
explicitly. Retracing our steps is indeed possible by making use of so-called root-separators $\hat k_n$, which 
play an important role in the theory of regular quantum graphs \cite{MF}. A separator $\hat k_n$ separates roots
number $n-1$ and number $n$ from each other. Because the theory of regular quantum graphs applies to $g^{(M)}(k)$,
we know the separators $\hat k^{(M)}_q$. To go one level up to $M-1$, we need to know the separators 
$\hat k_r^{(M-1)}$ which separate roots number $r-1$ and number $r$ from each other.
It turns out \cite{FPRE} that the extrema of $g^{(M-1)}(k)$ are root separators on the level $M-1$. But the location 
of the extrema of $g^{(M-1)}(k)$ are the zeros of $g^{(M)}(k)$, which we know. With the knowledge of the root separators
$\hat k_r^{(M-1)}$ we now compute the zeros $k_r^{(M-1)}$ of $g^{(M-1)}(k)$:
\begin{equation}
   k_r^{(M-1)}=\int_{\hat k_{r-1}^{(M-1)}}^
   {\hat k_r^{(M-1)}}\, k\, |g^{(M)}(k)|\,
   \delta(g^{(M-1)}(k))\, dk.
   \label{4}
\end{equation}
It is important to realize that (\ref{4}) is not just a formal solution, but yields $k_r^{(M-1)}$ explicitly, 
analytically, even numerically, if we wish, to any desired accuracy. This is so since both $g^{(M-1)}(k)$ and the
root separators $\hat k_r^{(M-1)}$ are known explicitly.
It is furthermore important to realize that the ``trick'' (\ref{4}) does not generally work for other quantum systems, 
even if their spectral equation $g$ is known. This is so because the root separators $\hat k_n$, known and explicitly 
constructable in the case of quantum graphs, are not in general known for other quantum systems. So for more general 
quantum (chaotic) systems the missing separators is indeed a hedge that, so far, protects them against explicit solution.

Our retracing process from levels $M$ to $M-1$ can now be continued until we reach level 0. This completes our proof that 
all scaling quantum graphs can be solved analytically in closed form.

\section{Acknowledgments}
The authors acknowledge financial support by the
National Science Foundation under Grant No. 9984075.

\end{document}